\documentclass[sigconf]{acmart}

\makeatletter                   
\def\mdseries@tt{m}             
\makeatother                    

\usepackage[draft=true]{minted}
\usepackage{graphicx}
\usepackage{mathrsfs}
\usepackage{stfloats}
\usepackage{color}
\hypersetup{
    colorlinks=true,
    linkcolor=blue,
    filecolor=red,      
    urlcolor=magenta,
    breaklinks=true,            
}

\AtBeginDocument{%
  \providecommand\BibTeX{{%
    \normalfont B\kern-0.5em{\scshape i\kern-0.25em b}\kern-0.8em\TeX}}}

\setcopyright{acmcopyright}
\copyrightyear{2019}
\acmYear{2019}

\acmConference[Anchorage '19]{The first workshop on Truth Discovery and Fact Checking: Theory and Practice, Anchorage, AK, USA}{August 05 2019}{Anchorage, AK, USA}
\acmBooktitle{Anchorage '19: The first workshop on Truth Discovery and Fact Checking: Theory and Practice, August 05 2019, Anchorage, AK, USA}

\begin{document}

\title{A Data Set of Internet Claims and Comparison of their Sentiments with Credibility}

\author{Amey Parundekar}
\affiliation{%
  \institution{Vellore Institute of Technology}
  \city{Chennai}
  \state{India}
  \postcode{600048}
}
\email{parundekaramey@gmail.com}

\author{Dr. Susan Elias}
\affiliation{%
  \institution{Vellore Institute of Technology}
  \city{Chennai}
  \state{India}
  \postcode{600048}}
\email{susan.elias@vit.ac.in}

\author{Dr. Ashwin Ashok}
\affiliation{%
  \institution{Georgia State University}
  \city{Atlanta}
  \country{USA}
}
\email{aashok@gsu.edu}

\renewcommand{\shortauthors}{Parundekar, et al.}

\begin{abstract}
In this modern era, communication has become faster and easier. This means fallacious information can spread as fast as reality. Considering the damage that fake news kindles on the psychology of people and the fact that such news proliferates faster than truth \cite{fasterspread}, we need to study the phenomenon that helps spread fake news. An unbiased data set that depends on reality for rating news is necessary to construct predictive models for its classification. This paper describes the methodology to create such a data set. We collect our data from snopes.com which is a fact checking organisation. Furthermore, we intend to create this data set not only for classification of the news but also to find patterns that reason the intent behind misinformation. We also formally define an \textit{Internet Claim}, its credibility, and \textit{sentiment} behind such a claim. We try to realize the relationship between the sentiment of a claim with its credibility. This relationship pours light on the bigger picture behind the propagation of misinformation. We pave the way for further research based on the methodology described in this paper to create the data set and usage of predictive modeling along with research based on psychology/mentality of people to understand why fake news spreads much faster than reality. 
\end{abstract}


\maketitle
\section{Introduction}

\subsection{Purpose}
Since the dawn of the internet, it has become increasingly easier to communicate information to anyone in this world at any time. Today, any piece of information travels faster over the world than it ever did. Advancements in technology have enabled us to reach an increasingly higher number of people in lower amounts of time. Although this proves to be a boon to society, it has several drawbacks too. It is paramount to broadcast critically essential information, related to food supplies and medical help, during a natural disaster and current technology does a perfect job at doing so. But, this technology also helps people spread fake news around the world that might at times critically affect the way people behave. Hence it has become necessary to sift between content over the internet to separate truth from the mendacious. An article recently published by Scientific American \cite{scientific_american} stated that, lately, social media has been the top source for news, for the majority of people in a country like the USA among others. 

In this paper we describe the development of a data set, needed for research, aimed at analyzing random claims made by people over various social media and their credit rating as rated by \textit{snopes.com}. It further aims to compare the sentiment of these claim with their credibility. This comparison helps us analyse the intent behind propagation of misinformation.   \textit{snopes.com} is one of the many fact-checking sources over the internet and we have used it to create our data set. The data set consists of internet claims, their ratings, sentimental analysis of the ratings, the origin of the claims and analysis of the claims in general. We not only take \textit{true} and \textit{false} as labels/ratings into consideration but also consider some other labels/ratings like \textit{mis-attribution} or \textit{mis-captioned} in this data set. This paper aims at discussing simple methods to extract such data over the internet in a time and cost effective way and finding a relation between the sentiment of such claims and either their truthfulness or falsehood.

\subsection{Challenges and Proposed Approach}
One of the major challenges in the classification of news is the identification of credible sources that act as reference classifiers. These, mainly, are fact-checking organizations. The verisimilitude of misinformation makes it very difficult for such classifiers and hence any predictive models to draw a line between reality and fiction. There are several factors involved, that need to be analyzed before we avow any piece of information as truth. One of the major factor, that often gets neglected, and thus pose a major challenge on current news classifiers is the \textbf{temporal}. A piece of information might hold at a particular time and yet at certain other times regarded fallacious. Another such important factor is the \textbf{sentiment} behind the news. A piece of information can either please everyone or infuriate them thereby causing overall conflict in the society. Hence, the real challenge faced by current day misinformation classifiers is the appropriate incorporation of these factors into their data sets and learning models.

For the collection of data that results in the creation of such a data set we propose an approach that first formally defines all things necessary for maintaining the authenticity of this data set. After doing so, we use a combination of web mining and natural language processing to finally generate the data set that pulls data from online fact-checking organizations. We also propose to incorporate temporal and sentimental factors in this data set. Hence instead of being a binary classification, we build a n-nary data set, where n can depend on the several factors that we account into the description of data. This way, we account for the temporal factor. Information, besides being \textbf{\textit{true}} and \textbf{\textit{false}} can also be \textbf{\textit{outdated}}, which means, in present time, the analysis of truth or fallacy of that piece of information has become irrelevant.

\section{Related Work}
There have been several efforts in the past to analyze fake news on social media. Most of these sources consider fake news detection as a binary classification problem. Particularly a paper on data mining perspective of fake news published by the Computer Science and Engineering department of the Arizona State University, Tempe \cite{asu_kdd} defines fake news as \textit{a news article that is intentionally and verifiable false}. They also define the prediction function for fake news detection as:
\[ 
\label{eqn:predict}
\mathcal{F(\textit{a})} =
\Big\{
  \begin{tabular}{ll}
  1, if \textit{a} is a piece of fake news,  \\
  0, otherwise.
  \end{tabular}
\]where \textit{a} is a news article and ${\mathcal{F}}$
is the prediction function that we want to learn. This definition seems quite apt for building a simple binary classifier but in further sections, we will build upon and extend this definition to support a multi-class classification model to supplement our data set. 

Several efforts have also been made in the past for the creation of data sets that complement research based on the detection of fake news. One such comprehensive data set is the CREDBANK \cite{CREDBANK} data set which is a big corpus of Tweets and their credit ratings as assessed by 30 Amazon Mechanical Turks. CREDBANK's creators thus created a good blend of manpower and computation to create their data set. A sample of their data set as provided on their GitHub page looks as shown in Table \ref{table:credbank}.

\begin{table}
    \caption{CREDBANK SAMPLE DATA}
    \begin{tabular}{*{4}{p{\dimexpr0.33\linewidth-5.5\tabcolsep\relax}}}
        \hline
    \textbf{topic\_key} & \textbf{topic\_terms} &  \textbf{Cred\_Ratings}  &
    \textbf{Reason}\\
        \hline\hline
        louis\_ebola\_...  &     [u'louis', u\'ebola\', u\'nurse\'] &  [\'1\', \'-1\', \'2\', \'-2\', \'0\', \'2\', \'0\',....] & [\'Nurses union describes the procedures taken by nurse who now has Ebola from treating a patient., .....]
    \end{tabular}
    \label{table:credbank}
\end{table}
 Here as we see, we have Tweets tokenized after removal of stopping words and presented in the table as tokens of keywords and not directly as tweets. In the Cred\_Ratings and Reason columns of the table, we see a list of ratings of credibility, rated on a scale from ${-2}$ to ${2}$ in the prior column and respective reason for that particular rating in the next. Each entry in this list of ratings and reasons is representing an Amazon Mechanical Turk, providing his/her own rating and their reason for that rating.
 Researchers at the Indiana University Observatory on Social Media have made several strides in the field of fake news detection as well. In doing so, they have launched several applications that study fake news. Hoaxy \cite{hoaxy} is a good example of their work. As their website describes, "Hoaxy is a search engine that shows users how stories from low-credibility sources spread on Twitter." 
 
\begin{figure*}
\centering
\includegraphics[width = \textwidth, height= \textheight, keepaspectratio]{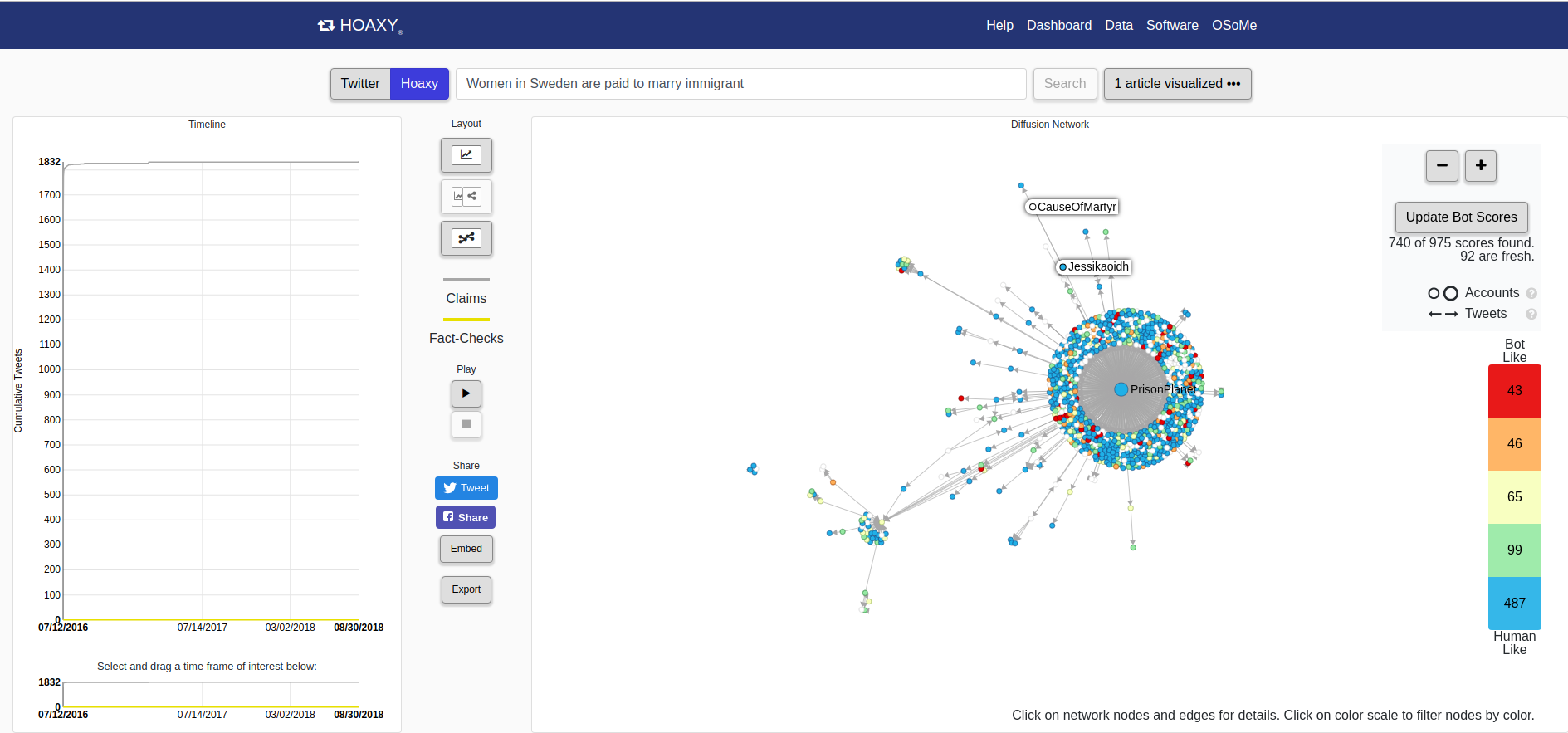}
\caption{Example of Hoaxy's network plot.}
\label{fig:hoaxy}
\end{figure*}

Figure \ref{fig:hoaxy}, shows a sample output of what Hoaxy is capable of. This is a plot for the search query: "Women in Sweden are paid to marry immigrants". 

Another study at MIT, analyzed all Tweets from 2006 to 2017 to find patterns among them. They stated that false news spreads significantly farther, faster, deeper, and more broadly than the truth. \cite{fasterspread}

\section{Background}
The studies in the previous section and their results, suggests that it is necessary to study the sentiment behind false claims to understand the bigger picture. Some people spread false news on purpose but some people seem to spread such news unknowingly and without verification. The sentiment of a claim sheds a light on why people knowingly/unknowingly end up making/sharing false claims faster, deeper and farther than true claims. We start by formally defining an \textit{Internet Claim}, \textit{Fact-Checking function} for a claim and \textit{Sentiment} of a claim.

\subsection{Internet Claim}
An \textbf{Internet Claim} is any piece of information published/written or has any other form of presence over the internet visible to all on any media by a person or an entity that might be true or false.

Only after proper examination of such a claim and it's fact-checking, should we make any conclusion about the authenticity of such a claim or form an opinion about it. Notice that the word "should" here is very important as it highlights the fact that we "can" form opinions but "should not" do so without verification. This might be a restatement but it is very essential for understanding the observed phenomenon of faster false news travel. 

\subsection{Fact-Checking Function}
\textbf{Fact-Checking Function} for an Internet Claim is a function that rates the credibility of the claim. Notice that this is \textbf{\textit{not}} a \textbf{predictive function} like the one given in Equation \ref{eqn:predict} but rather an \textbf{assertive function} that rates the credibility of an Internet Claim without any sort of learning or prediction. The structure and output of this function is based possibly on ground truth and reality which is what we desire but it's sometimes also based on belief. One possible but not limiting mathematical model of this function is:

\[ 
\label{eqn:assert}
\mathcal{S(\textit{a})} =
\Big\{
  \begin{tabular}{ll}
  False, if \textit{a} is a piece of fake news,  \\
  True, otherwise.
  \end{tabular}
\]where \textit{a} is a news article and ${\mathcal{S}}$ is an assertion function that rates the credibility of claims.

Wait, maths and beliefs? What are we talking about here? Right? Well, to give an example, "beliefs" are very much like the postulates we see and agree to be true without any proof in Euclidean Geometry. It is important to notice that, instead of thinking of this as a mathematical function we need to think of the Fact-Checking Function as a fact-checking organization like snopes.com. Of course, goes without saying that a Fact-Checking Function should, assert a truthful output, failing to which we would not be sure if our credibility ratings for a claim are correct or not. This, in fact, means that we need to assess the credibility of a Fact-Checking Function, which for our examples means that we need to verify that snopes.com is a credible source and does not provide us with wrong ratings for news.

As we observe, we face a challenge where we have to apply another Fact-Checking Function on the initial Fact-Checking Function to rate the credibility of the Fact-Checking Function itself so that we know for sure that our ratings are correct or not. Notice that of course, the two Fact-Checking Functions here should be different from each other and not the same. 
\[ 
\mathcal{S_{\mathbf{2}}(S_{\mathbf{1}}(\textit{a}))} =
\Big\{
  \begin{tabular}{ll}
  False, if $\mathcal{S_{\mathbf{1}}}$ is source of false ratings,\\
  True, otherwise.
  \end{tabular}
\]where \textit{a} is a news article and ${S_{\mathbf{1}}}$ is the initial credibility assertion function that rates the credibility of the claims and ${S_{\mathbf{2}}}$ is the credibility assertion function for the initial function.
To overcome this conundrum, we define something called the Event function. 

\subsection{Event Function}
Event Function for an Internet Claim is the function that is based on reality and not on belief. It tells us what actually happened. It gives a binary output of either True if the event described in a claim actually happened or False if the event described in a claim never happened in this reality. Notice that such a binary behavior is not necessarily expected by a Fact-Checking Function. A Fact-Checking Function can be ternary or have even higher orders. (with example states like True, False, Mostly True, Mostly False, Mis-captioned, etc.).
\[ 
\mathcal{E(\textit{a})} =
\Bigg\{
  \begin{tabular}{ll}
  \textbf{True}, if \textit{a} is a is piece of claim that describes\\ \quad\qquad an event that    actually occurred in reality,  \\
  \textbf{False}, otherwise.
  \end{tabular}
\]where \textit{a} is a news article and ${\mathcal{E}}$ is an assertion function that rates the credibility of claims.

We thus say that any Fact-Checking Function that is derived from/based upon the Event Function correctly assess the credibility of an Internet Claim and then that function and any Fact-Checking Function(s) derived from it can be considered as credible rating sources. This is like saying that a representative from snopes.com actually went on to look for physical details of an event and found conclusive proofs in favor of the event or against it or they derived their results from some other fact-checking organizations which found such conclusive proofs which made both of their information credible. \par
After sifting through all resources available online, to collect data related to internet claims, we chose snopes.com  as it turned out to be a very good source to collect the data for several interesting reasons. On observing snopes.com's Twitter account through the Twitter API \cite{TwitterPlatform}, it was found that the account had tweets of \textit{Internet Claims}, with every tweet containing a link to their website, which had that claim from the internet analyzed on being True, False, Mis-captioned, etc. along with proper reasoning and comparison with the ground truth to verify their rating for the claim provided as "Origin". This makes snopes.com a Fact-Checking Function that is dependent directly on the Event Function hence making their credibility ratings credible.

\section{Sentiment of a Claim}
The sentiment of an Internet Claim is the emotional bias of the claim. It can mainly be classified into Positive, Negative and Neutral. To put it in common terms, the Sentiment of a claim is how you feel about the claim when you read it. If it sparks off anger, then the claim has a negative sentiment. Whereas if it fosters joy or courage or happiness, the claim has a positive sentiment. And if neither happens then the claim has a neutral sentiment. For a article \textit{a}, the \textit{sentiment} can be defined as:

\[ 
\mathcal{B(\textit{a})}
\label{eqn:sentiment}
\Bigg\{
  \begin{tabular}{lll}
  ${>}$ 0, if positive\\
  ${=}$ 0, if neutral, \\
  ${<}$ 0, if negative
  \end{tabular}
\]

\begin{figure*}
\centering
\includegraphics[width = \textwidth,height = \textheight, keepaspectratio]{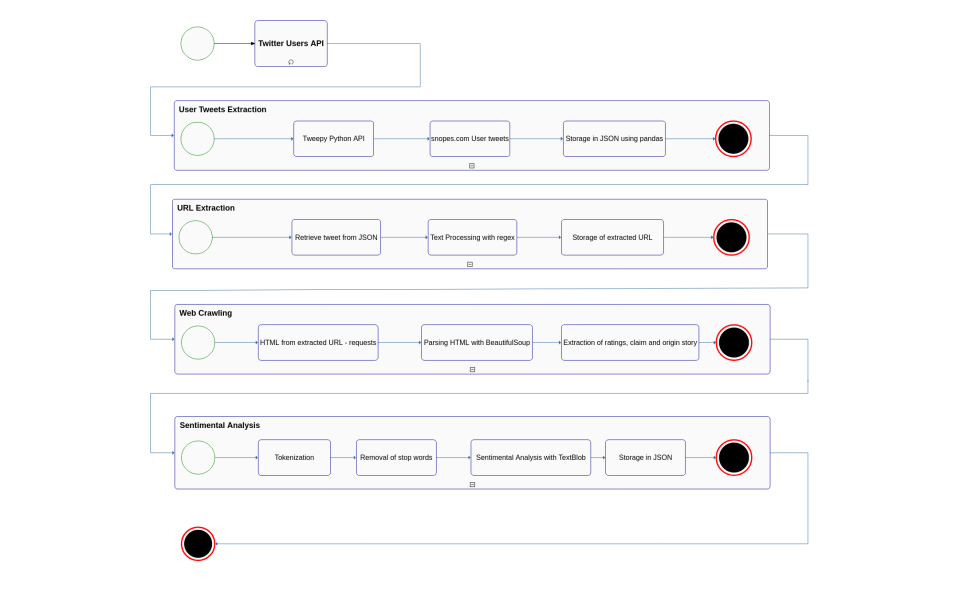}
\caption{Process flow of data set creation}
\label{fig:processflow}
\end{figure*}

\subsection{Getting Data}
The process followed by us for creation of the data set is given in the Process Flow diagram \ref{fig:processflow}.
The code to get data with Twitter API was written in python and inspired by Vincent Russo's GitHub repository \cite{LucidProgramming} on working with tweepy \cite{tweepy} which is a python library for the Twitter Developer's API. The code was written to collect user tweets of snopes.com's Twitter account for the reason as mentioned in section 3.3. According to Twitter's API limitation's one can extract 15 pages worth tweets with 200 tweets per page from a user's profile. Keeping this in mind, tweets were extracted from snopes.com's profile multiple times with enough time gap between two collections, making sure that we don't collect already collected tweets. These tweets were stored in 15 separate files for each page, with 200 tweets in each file in JSON format. Sample tweet data is as follows:

\begin{verbatim}
"8": {
        "tweets": "Was Bill O'Reilly 
                   found dead at his 
                   Long Island home? 
          https:\/\/t.co\/SGwagACMbW 
         https:\/\/t.co\/Ppx1FhJeMm",
        "id": 1075020507186126853,
        "len": 101,
        "date": 1545139836000,
        "source": "AgoraPulse Manager",
        "likes": 4,
        "retweets": 2,
        "time": 1545139836000,
        "geo": null,
        "sentiment": -1,
        "token_list": [
            "Was",
            "Bill",
            "O",
            "Reilly",
            "found",
            "dead",
            "Long",
            "Island",
            "home"
        ]
    }
\end{verbatim}
This is the eighth tweet from the first page's output file. Among the several available parameters provided by twitter \cite{twitterparams} the one seen in the above example were used. Pandas \cite{pandas} was used to deal with data frames used to handle the above data.

\subsection{URL Extraction}
As it is observed, the "tweets" section which contains the text from the actual tweet, contains a URL which maps to the corresponding post on snopes.com for the claim mentioned in that tweet. 
Hence, "\textit{https:\/\/t.co\/SGwagACMbW}" in fact maps to "\textit{https://www.snopes.com/fact-check/bill-oreilly-found-dead/}". 

\begin{figure}[h]
\centering
\includegraphics[width = 8.55cm]{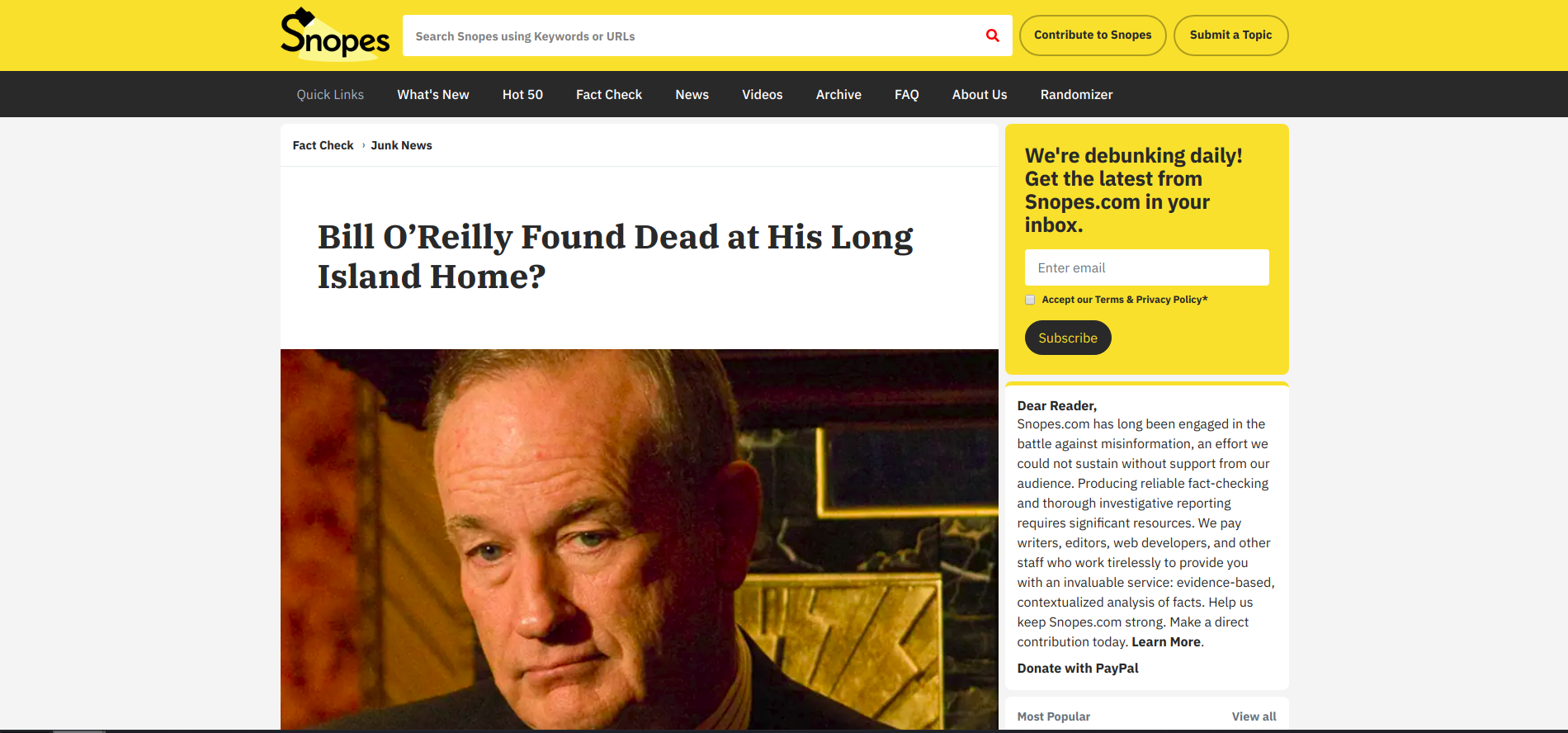}
\caption{snopes.com/fact-check/bill-oreilly-found-dead/}
\label{fig:snopes}
\end{figure}

\begin{figure}[h]
\centering
\includegraphics[width = 8.55cm]{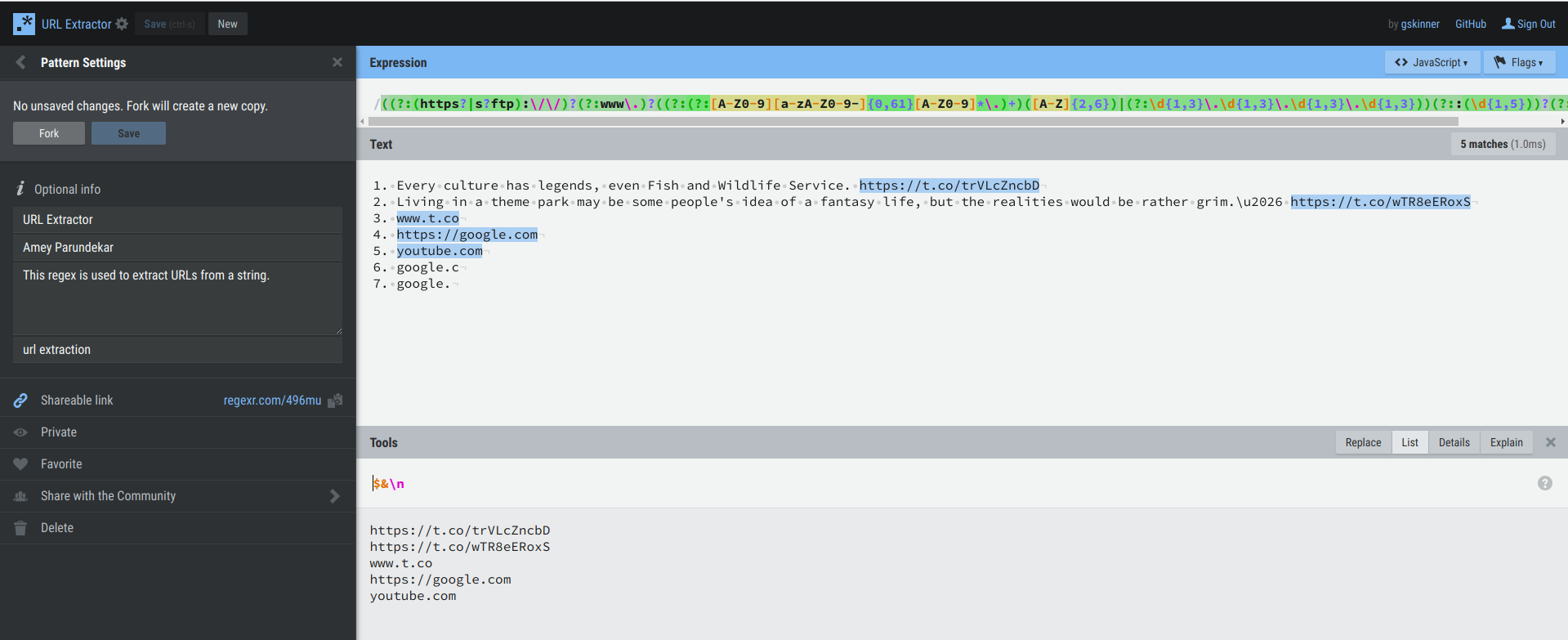}
\caption{URL Extractor}
\label{fig:regexr}
\end{figure}

The page in Figure \ref{fig:snopes} contains 3 main sections which are: \textbf{\textit{Claim}}, \textbf{\textit{Rating}} and \textbf{\textit{Origin}}. The claim section contains the Internet Claim, the rating section contains the rating of this claim and the origin section gives an explanation about the origin of this claim. So our job amounts to extracting these URLs from the tweets of snopes.com's Twitter account. The most straight forward and elegant solution to extract these URLs is via the use of regular expressions. We used the following regular expression:
\begin{verbatim}
((?:(https?|s?ftp):\/\/)?(?:www\.)?
((?:(?:[A-Z0-9][a-zA-Z0-9-]{0,61}
[A-Z0-9]*\.)+)([A-Z]{2,6})|(?:\d
{1,3}\.\d{1,3}\.\d{1,3}\.\d{1,3}))
(?::(\d{1,5}))?(?:(\/\S+)*))
\end{verbatim}
This regular expression extracts all URLs from a text which contain http, https or ftp. It takes care of presence or absence of \textbf{www}, and the length of the URL. For further explanation on how this regular expressions works, one can have a look at the regex link \cite{regexr} that we created. A few examples of how the regular expression works is given in Figure \ref{fig:regexr}.

\subsection{Web Crawling}
After extraction of the URLs from the snopes.com's profile's tweets, we get the HTML response of their web pages which have claims and ratings. Our work, then, amounts to finding the sections of the page which provide us with the claim and the rating information for that claim. After obtaining the ratings for the claim, it is stored back into the tweet's JSON data. We make use of BeautifulSoup \cite{BeautifulSoup} which is a python library to parse HTML and XML content, to extract this data from the URL's HTML response. A sample of the rated tweets or claims is shown next.

\begin{verbatim}
     "8": {
    "origin-html": "[<div class=\"post-
    body-card post-card card\">\\n<h3 
    class=\"card-header\"> Origin</h3>
    \\n<div class=\"card-body\">\\n<p>
    On 21 May 2017, the Daily USA 
    Update web site published
    an article purporting to reveal 
    \\u201cmore details about
    the sad death\\u201d of 
    former Fox News anchor
    Bill O\\u2019Reilly:
    </p>\\n<blockquote><p>
    The Islip Coroner\\u2019s 
    Office stated that
    last night,...]"
    "token_list": [
      "Was" ,
      "Bill" ,
      "O" ,
      "Reilly" ,
      "found" ,
      "dead" ,
      "Long" ,
      "Island" ,
      "home"
    ] ,
    "source": "AgoraPulse Manager" ,
    "len": 101 ,
    "claim-html": "[<p class=\"claim\">
                    Former Fox News host 
                    Bill O'Reilly 
                    was found dead on 
                    Long Island.</p>]" ,
    "date": 1545139836000 ,
    "rating-html": "[<span class=
    \"rating-name 
    rating-label-false\">False</span>]" ,
    "likes": 4 ,
    "time": 1545139836000 ,
    "tweets": "Was Bill O'Reilly 
    found dead at his 
    Long Island home? 
    https://t.co/SGwagACMbW 
    https://t.co/Ppx1FhJeMm" ,
    "geo": null ,
    "id": 1075020507186126853 ,
    "retweets": 2
  } ,
\end{verbatim}

\begin{figure*}
  \includegraphics[width=\textwidth,height=\textheight, keepaspectratio]{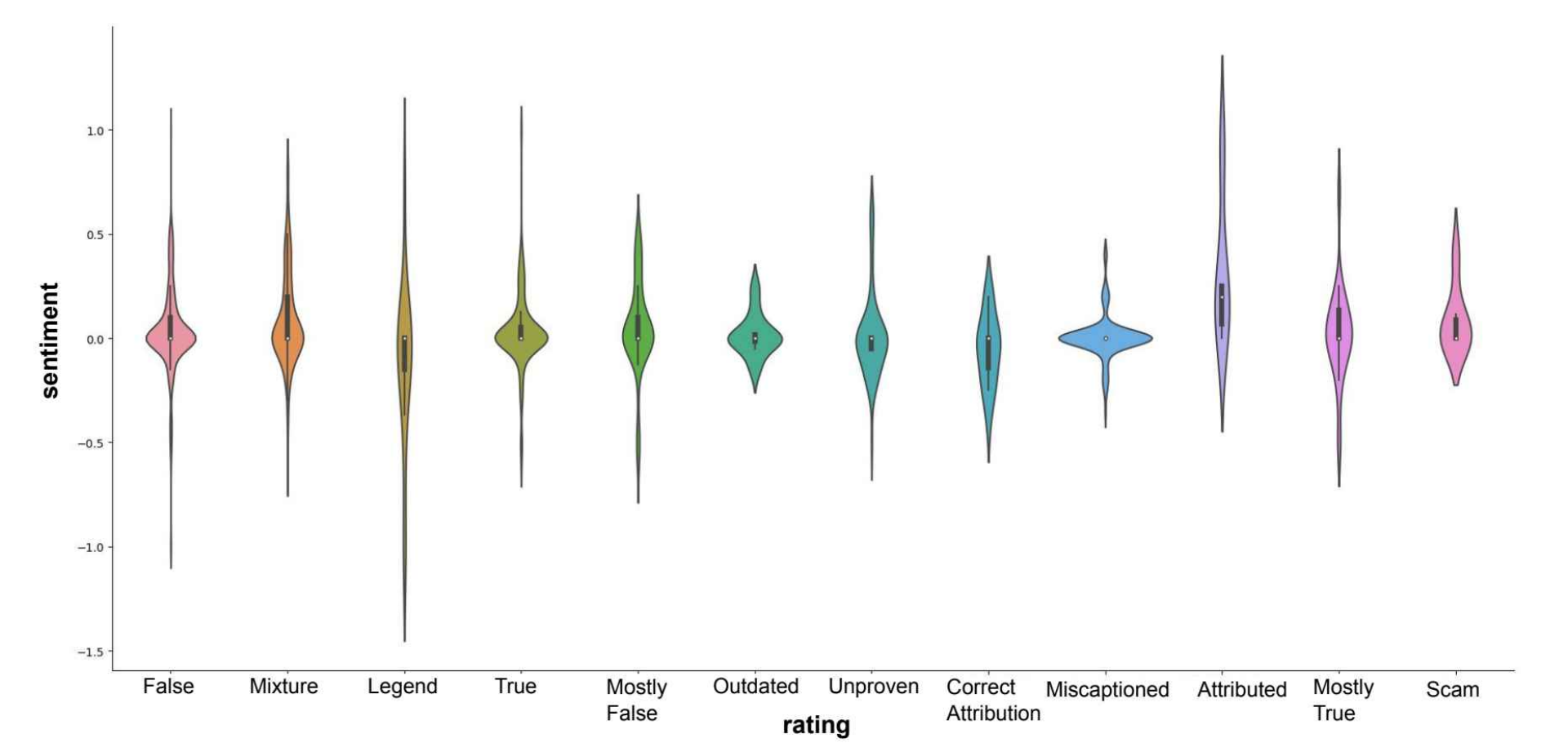}
  \caption{Violin plot of Sentiment vs Ratings}
  \label{fig:mainviolin}
\end{figure*}

We have three  new additions to the initial JSON data. \textit{"origin-html"}, \textit{"rating-html"} and \textit{"claim-html"}. Hence we were successfully able to rate the claim and also find its origin story!

\subsection{Sentiment Analysis of the Rated Claims.}
TextBlob \cite{textblob} is a python library used for text processing. It contains APIs that take in a string of text and return the overall sentiment expressed by the words in that string. It rates the sentiment between $-1$ to $1$ and follows the logic we developed for sentiment in Section 3.3. Hence we get a normalized sentiment analysis of our claims via this python library. 

\subsection{Assembly of Data Set}
We finally gathered together all the claims, their ratings, their origin story, the sentiment of these claims and nicely bundle them into a comma separated values. In doing so we get rid of all the claims that we couldn't find any rating information for. Example of the csv output and how our data set looks like is given below.

\begin{verbatim}
[Former Fox News 
host Bill O'Reilly was found dead on
Long Island.], False, 
-0.083333333333333, 
[On 21 May 2017, 
the Daily USA Update web site
published an article purporting to
reveal “more details about the sad 
death” of former Fox News anchor 
Bill O’Reilly:...]
\end{verbatim}
The first value is the claim, second value - "False" is the rating of the claim, the third value is the sentiment of the claim. And as we can see, it has a negative sentiment and the fourth value is the origin story. You can find this data set at the GitHub page \cite{dataset} and have a glimpse of how it looks like in Table \ref{table:iclaimnet}. \par
Finding that many tweets that snopes.com's Twitter account posted contained fact-checked internet claims and a link to their site containing detailed analysis of these internet claims was quite serendipitous. Without making this trivial discovery, we would not have been able to generate the data set. Notice that Twitter acted only as a passive medium that allowed us to mine the fact-checked internet claims from snopes.com. We did not in any way fact-check tweets but in fact used a "hack" we found in one of the fact-checking organisation's twitter account to our benefit. The data collected for the data set generated in the process described here, was collected from the most recent tweets of snopes.com's twitter account, which resulted in collection of the corresponding internet claims. Since the order in which snopes.com posted the internet claims on twitter was random, in the sense that even though it may or may not have been temporally serial, it formed no traceable patterns among itself, and every internet claim collected in such way was different from every other, the data collected was unrelated, mutually exclusive and hence formed a random pool of internet claims.

\begin{table}
    \caption{iClaimNet DATASET\\www.github.com/the-lost-explorer/iClaimNet}
    \begin{tabular}{*{4}{p{\dimexpr0.33\linewidth-5.5\tabcolsep\relax}}}
        \hline
    \textbf{claim} & \textbf{rating} &  \textbf{sentiment}  &
    \textbf{origin}\\
        \hline\hline
        [Former Fox News host Bill O'Reilly was found dead on Long Island.] &  FALSE &  -0.083333333
 & [On 21 May 2017,the Daily USA Update web site published an article purporting to reveal â more details about the sad death of former Fox News anchor Bill O'Reilly..]
    \end{tabular}
    \label{table:iclaimnet}
\end{table}

\section{Observations}

With our \textit{comma separated values} of claims, ratings and their sentiments, one can start visualizing what is the sentimental trend behind a particular rating. We plot the sentimental value of a claim to its rating as seen in Figure 4. We have a \textit{Violin Plot} of the ratings vs sentiment for every claim. On the X axis we have 12 different rating types---\textit{\textbf{True, False, Mostly True, Mostly False, Outdated, Mis-captioned, Mis-attributed, Unproven, Mixture, Legend, Scam, Correct Attribution}}.
A description of what each rating represents is given on snopes.com's rating page \cite{snopesratings}. We demonstrate our data using a violin plot. A violin plot is a method of plotting numeric data. It is similar to a box plot, with the addition of a rotated kernel density plot on each side. Violin plots are similar to box plots, except that they also show the probability density of the data at different values, usually smoothed by a kernel density estimator. The violin plot of Figure \ref{fig:mainviolin} depicts the standard distribution, inter-quartile range and median of the sentiment score for each rating. 

\begin{figure*}
\centering
\includegraphics[width = \textwidth, height = \textheight, keepaspectratio]{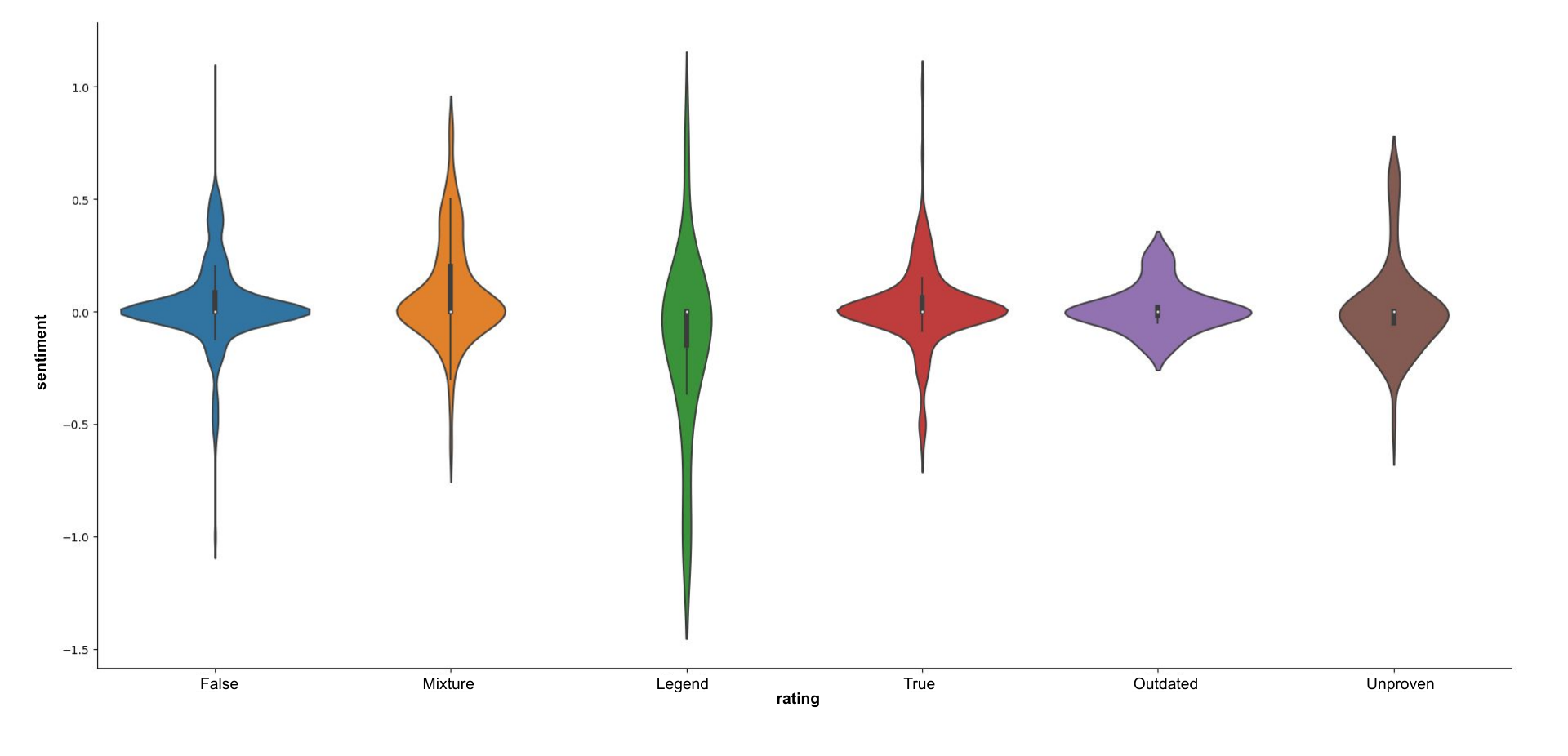}
\caption{Violin Plot of Clustered Ratings}
\label{fig:cviolin}
\end{figure*}

Here we can further cluster "False", "Mostly False", "Mis-attributed", "Mis-captioned" and "Scam" claims under a single category and "True", "Mostly True" and "Correct Attribution" under another single category due to a similarity in the meaning they express.
The violin plot after such clustering is given in Figure \ref{fig:cviolin}. 

The statics among the 1669 analyzed claims, is as follows:
\begin{verbatim}
Total False Claims: 495
Total True Claims: 160
False Positives: 306
False Negatives: 189
True Positives: 99
True Negatives: 61
\end{verbatim}

\textbf{Observation 1:}
Upon discussion in the section on assembly of data set we understood that the data set forms a random pool of internet claims and in such a random sample of about 1600 internet claims, the number of false claims is much higher than the number of true claims.

\textbf{Observation 2:}
Moreover ${38.18}\%$ of the false claims had a negative sentiment whereas $38.12\%$ of the true claims had a negative sentiment. These two figures are almost the same. 

\textbf{Observation 3:}
When we look at the distribution graph of the sentiments(see Figure \ref{fig:cviolin}) we observe that claims that have a very high negative sentiment are false whereas, the claims that have a very high positive sentiment can be both true or false. This can also be verified by the figures. When we compare the false and true claims, we find only 4 claims with a sentiment lower than -0.6 and all of them are false claims. As opposed to that, we find only 5 claims with a sentiment above 0.6 and they are either true or false.

Generation of more data may pour more light into this observation. We have only just started observing patterns that relate the credibility of Internet Claims to their sentiments. This will help us understand the reason why people choose to spread fallacious claims. Our observations, based on limited data, suggest that highly negative claims which burgeon negativity among masses have a tendency of being fallacious. This tell us that false claims are made to bring about a sense of negativity among masses.
\section{Conclusion}
The research aimed at fake news analysis, which has data labeled with ratings of news/claims/tweets/internet articles, to perform certain machine learning predictions, needs to verify the credibility of those ratings before proceeding with the prediction. This paper formally defines the conditions necessary for assessing the credibility of the ratings of internet claims. 
This paper also describes a methodology used to create a data set for analysis of internet claims or fake news analysis and any predictive research aimed at the detection of fake news online. In the process of creating this data set, we also made observations about the sentiment behind these claims and compared the true positives with the false positives and true negatives with false negatives. We observed that both true and false claims had about 31\% of the claims positive and the rest, negative. This means we cannot generally comment on the sentiment of a false or a true claim. But we can conclude that, if we have a highly negative claim, it has a tendency of being false. We also conclude after looking at the numbers that there were more false news articles/false claims than true news articles/true claims. In any given random sample, there is more fake news than reality. 
\section{Future Work}
This research needs to be further expanded to use techniques for data collection as described here to extract an analyze more internet claims to create a large scale corpus of internet claims. We also need to use the claims from this data set separately for every social media platform, to analyze the time series data of such claims spread by people on those media. Only on proper analysis of such time series data can we come up with conclusive machine learning algorithms to predict if any given claim is true or false.
Furthermore, research aimed at understanding the psychology of people is necessary to understand why fake news travels much faster than truth and why there is more fake news and false claims over the internet than truth. 

%
%
%

\nocite{*}
\bibliographystyle{ACM-Reference-Format}
\bibliography{refs}


\begin{thebibliography}{00}


\ifx \showCODEN    \undefined \def \showCODEN     #1{\unskip}     \fi
\ifx \showDOI      \undefined \def \showDOI       #1{#1}\fi
\ifx \showISBNx    \undefined \def \showISBNx     #1{\unskip}     \fi
\ifx \showISBNxiii \undefined \def \showISBNxiii  #1{\unskip}     \fi
\ifx \showISSN     \undefined \def \showISSN      #1{\unskip}     \fi
\ifx \showLCCN     \undefined \def \showLCCN      #1{\unskip}     \fi
\ifx \shownote     \undefined \def \shownote      #1{#1}          \fi
\ifx \showarticletitle \undefined \def \showarticletitle #1{#1}   \fi
\ifx \showURL      \undefined \def \showURL       {\relax}        \fi
\providecommand\bibfield[2]{#2}
\providecommand\bibinfo[2]{#2}
\providecommand\natexlab[1]{#1}
\providecommand\showeprint[2][]{arXiv:#2}

\bibitem[\protect\citeauthoryear{American}{American}{2018}]%
        {scientific_american}
\bibfield{author}{\bibinfo{person}{Scientific American}.}
  \bibinfo{year}{2018}\natexlab{}.
\newblock \bibinfo{title}{Biases Make People Vulnerable to Misinformation
  Spread by Social Media}.
\newblock
  \bibinfo{howpublished}{\url{https://www.scientificamerican.com/article/biases-make-people-vulnerable-to-misinformation-spread-by-social-media/}}.
    (\bibinfo{date}{Jun} \bibinfo{year}{2018}).
\newblock


\bibitem[\protect\citeauthoryear{Crummy.com}{Crummy.com}{2019}]%
        {BeautifulSoup}
\bibfield{author}{\bibinfo{person}{Crummy.com}.}
  \bibinfo{year}{2019}\natexlab{}.
\newblock \bibinfo{title}{Beautiful Soup-a Python library for pulling data out
  of HTML and XML files.}
\newblock \bibinfo{howpublished}{\url{
  https://www.crummy.com/software/BeautifulSoup/bs4/doc/ }}.
  (\bibinfo{year}{2019}).
\newblock


\bibitem[\protect\citeauthoryear{Mitra and Gilbert}{Mitra and Gilbert}{2015}]%
        {CREDBANK}
\bibfield{author}{\bibinfo{person}{Tanushree Mitra} {and} \bibinfo{person}{Eric
  Gilbert}.} \bibinfo{year}{2015}\natexlab{}.
\newblock \bibinfo{title}{CREDBANK: A Large-scale Social Media Corpus With
  Associated Credibility Annotations}.
\newblock \bibinfo{howpublished}{\url{
  https://github.com/compsocial/CREDBANK-data }}.   (\bibinfo{year}{2015}).
\newblock


\bibitem[\protect\citeauthoryear{Observatory~on Social~Media}{Observatory~on
  Social~Media}{2016}]%
        {hoaxy}
\bibfield{author}{\bibinfo{person}{Indiana~University Observatory~on
  Social~Media}.} \bibinfo{year}{2016}\natexlab{}.
\newblock \bibinfo{title}{Visualize the spread of claims and fact checking.}
\newblock \bibinfo{howpublished}{\url{ https://hoaxy.iuni.iu.edu/ }}.
  (\bibinfo{year}{2016}).
\newblock


\bibitem[\protect\citeauthoryear{Pandas.pydata.org}{Pandas.pydata.org}{2019}]%
        {pandas}
\bibfield{author}{\bibinfo{person}{Pandas.pydata.org}.}
  \bibinfo{year}{2019}\natexlab{}.
\newblock \bibinfo{title}{Pandas - Python Data Analysis Library}.
\newblock \bibinfo{howpublished}{\url{ https://pandas.pydata.org }}.
  (\bibinfo{year}{2019}).
\newblock


\bibitem[\protect\citeauthoryear{Parundekar}{Parundekar}{2019a}]%
        {dataset}
\bibfield{author}{\bibinfo{person}{Amey Parundekar}.}
  \bibinfo{year}{2019}\natexlab{a}.
\newblock \bibinfo{title}{Data set to accompany Truth Discovery and Fact
  Checking: Theory and Practice SIGKDD 2019 Workshop, August 5th, Anchorage,
  Alaska paper 'A Data Set of Internet Claims and Comparison of their
  Sentiments with Credibility'}.
\newblock \bibinfo{howpublished}{\url{
  https://github.com/the-lost-explorer/iClaimNet }}.   (\bibinfo{year}{2019}).
\newblock


\bibitem[\protect\citeauthoryear{Parundekar}{Parundekar}{2019b}]%
        {regexr}
\bibfield{author}{\bibinfo{person}{Amey Parundekar}.}
  \bibinfo{year}{2019}\natexlab{b}.
\newblock \bibinfo{title}{Regular Expression for URL Extraction.}
\newblock \bibinfo{howpublished}{\url{ https://regexr.com/496mu }}.
  (\bibinfo{year}{2019}).
\newblock


\bibitem[\protect\citeauthoryear{Russo}{Russo}{2017}]%
        {LucidProgramming}
\bibfield{author}{\bibinfo{person}{Vincent Russo}.}
  \bibinfo{year}{2017}\natexlab{}.
\newblock \bibinfo{title}{Vincent Russo's GitHub repository on tweepy API.}
\newblock \bibinfo{howpublished}{\url{
  https://github.com/vprusso/youtube\_tutorials/blob/master/twitter
  \_python/part\_1\_streaming\_tweets/tweepy\_streamer.py }}.
  (\bibinfo{year}{2017}).
\newblock


\bibitem[\protect\citeauthoryear{Shu, Sliva, Wang, Tang, and Liu}{Shu
  et~al\mbox{.}}{2016}]%
        {asu_kdd}
\bibfield{author}{\bibinfo{person}{Kai Shu}, \bibinfo{person}{Amy Sliva},
  \bibinfo{person}{Suhang Wang}, \bibinfo{person}{Jiliang Tang}, {and}
  \bibinfo{person}{Huan Liu}.} \bibinfo{year}{2016}\natexlab{}.
\newblock \bibinfo{title}{Fake News Detection on Social Media: A Data Mining
  Perspective}.
\newblock \bibinfo{howpublished}{\url{
  https://www.kdd.org/exploration\_files/19-1-Article2.pdf}}.
  (\bibinfo{year}{2016}).
\newblock


\bibitem[\protect\citeauthoryear{Snopes.com}{Snopes.com}{2019}]%
        {snopesratings}
\bibfield{author}{\bibinfo{person}{Snopes.com}.}
  \bibinfo{year}{2019}\natexlab{}.
\newblock \bibinfo{title}{Comprehensive list of the ratings snopes.com uses and
  their definitions}.
\newblock \bibinfo{howpublished}{\url{
  https://www.snopes.com/fact-check-ratings/ }}.   (\bibinfo{year}{2019}).
\newblock


\bibitem[\protect\citeauthoryear{Soroush~Vosoughi}{Soroush~Vosoughi}{2018}]%
        {fasterspread}
\bibfield{author}{\bibinfo{person}{Sinan~Aral Soroush~Vosoughi, Deb~Roy}.}
  \bibinfo{year}{2018}\natexlab{}.
\newblock \bibinfo{title}{The spread of true and false news online}.
\newblock \bibinfo{howpublished}{\url{
  http://science.sciencemag.org/content/359/6380/1146/tab-pdf }}.
  (\bibinfo{date}{Mar} \bibinfo{year}{2018}).
\newblock


\bibitem[\protect\citeauthoryear{TextBlob.com}{TextBlob.com}{2019}]%
        {textblob}
\bibfield{author}{\bibinfo{person}{TextBlob.com}.}
  \bibinfo{year}{2019}\natexlab{}.
\newblock \bibinfo{title}{TextBlob: Simplified Text Processing}.
\newblock \bibinfo{howpublished}{\url{ https://textblob.readthedocs.io/en/dev/
  }}.   (\bibinfo{year}{2019}).
\newblock


\bibitem[\protect\citeauthoryear{Tweepy.org}{Tweepy.org}{2019}]%
        {tweepy}
\bibfield{author}{\bibinfo{person}{Tweepy.org}.}
  \bibinfo{year}{2019}\natexlab{}.
\newblock \bibinfo{title}{An easy-to-use Python library for accessing the
  Twitter API.}
\newblock \bibinfo{howpublished}{\url{ https://www.tweepy.org }}.
  (\bibinfo{year}{2019}).
\newblock


\bibitem[\protect\citeauthoryear{Twitter}{Twitter}{2019a}]%
        {twitterparams}
\bibfield{author}{\bibinfo{person}{Twitter}.} \bibinfo{year}{2019}\natexlab{a}.
\newblock \bibinfo{title}{Twitter API parameters.}
\newblock \bibinfo{howpublished}{\url{
  https://developer.twitter.com/en/docs/tweets/timelines/api-reference/get-statuses-user\_timeline
  }}.   (\bibinfo{year}{2019}).
\newblock


\bibitem[\protect\citeauthoryear{Twitter}{Twitter}{2019b}]%
        {TwitterPlatform}
\bibfield{author}{\bibinfo{person}{Twitter}.} \bibinfo{year}{2019}\natexlab{b}.
\newblock \bibinfo{title}{Twitter's Platforms for developers.}
\newblock \bibinfo{howpublished}{\url{
  https://developer.twitter.com/en/docs/basics/getting-started }}.
  (\bibinfo{year}{2019}).
\newblock


\end{thebibliography}
\end{document}